\documentclass[sigconf]{acmart}
\AtBeginDocument{%
  \providecommand\BibTeX{{%
    \normalfont B\kern-0.5em{\scshape i\kern-0.25em b}\kern-0.8em\TeX}}}

\copyrightyear{2026}
\acmYear{2026}
\setcopyright{acmlicensed}\acmConference[UIST '26]{ACM Symposium on User Interface Software and Technology}{Nov 2--5, 2026}{Detroit, MI, USA}
\acmDOI{XXXXXXX.XXXXXXX}

\usepackage{color,soul}
\usepackage{booktabs} 
\usepackage{subfig} 
\usepackage{graphicx}

\usepackage[utf8]{inputenc}
\usepackage[T1]{fontenc}

\usepackage{siunitx}
\newcommand{\hide}[1]{}

\sethlcolor{yellow}
\ifdefined\scifihidecomments
    \newcommand{\cz}[1] {}
    \newcommand{\hyunc}[1] {} 
    \newcommand{\yax}[1] {} 
    \newcommand{\sony}[1] {} 
    \newcommand{\ReviewerFeedback}[1] {}  
    \newcommand{\fran}[1] {}
    \newcommand{\rz}[1]{}
    \newcommand{\lw}[1] {}
    \newcommand{\mose}[1] {} 
    \newcommand{\ke}[1] {} 

\else
    \definecolor{burntorange}{rgb}{0.8, 0.33, 0.0}
    \definecolor{cadmiumgreen}{rgb}{0.0, 0.42, 0.24}
    \definecolor{cobalt}{rgb}{0.0, 0.28, 0.67}
    \definecolor{amber}{rgb}{1.0, 0.75, 0.0}
    \definecolor{fashionfuchsia}{rgb}{0.96, 0.0, 0.63}
    \definecolor{brightcerulean}{rgb}{0.11, 0.67, 0.84}
    \definecolor{frenchblue}{rgb}{0.0, 0.45, 0.73}
    \definecolor{darkslateblue}{rgb}{0.28, 0.24, 0.55}
    \definecolor{cerulean}{rgb}{0.0, 0.48, 0.65}
    \definecolor{darkpastelgreen}{rgb}{0.01, 0.75, 0.24}

    \newcommand{\cz}[1] { \textcolor{red}{[\hl{cheng:} {#1}}]}
    \newcommand{\hyunc}[1] { \textcolor{burntorange}{[{\hl{hyunc:}} {#1}}]}
    \newcommand{\yax}[1] { \textcolor{magenta}{[{\hl{yaxuan:}} {#1}]}}
    
    \newcommand{\sony}[1] { \textcolor{blue}{[{\hl{songyun:}} {#1}]}}
    \newcommand{\ReviewerFeedback}[1] { \textcolor{brightcerulean}{[{Reviewer Feedback:} {#1}}]}
    \newcommand{\fran}[1]{\textcolor{burntorange}{[{francois:}{#1}}]}
    \newcommand{\rz}[1]{\textcolor{teal}{[{Ruidong: }{#1}]}}
    \newcommand{\lw}[1]{\textcolor{fashionfuchsia}{[{liuwei:}{#1}}]}
    \newcommand{\mose}[1]{\textcolor{burntorange}{[{mose:}{#1}}]}
    \newcommand{\ke}[1] { \textcolor{red!55!yellow}{[{Ke:} {#1}}]}

    \definecolor{CAT-comment}{rgb}{0.95, 0.2, 0.8}
    \definecolor{GL-comment}{rgb}{0.0, 0.54, 0.8}

\fi

\ifdefined\scifiblind
    \newcommand{\blind}[1]{[omitted for blind review]}
\else
    \newcommand{\blind}[1]{#1} 
\fi

\begin{document}

\title[]{KnitID: Machine-Knitted RFID Antennas for Battery-Free Authentication, Localization and Interaction}






\author{Weiye Xu}
\email{xuwy24@uw.edu}
\affiliation{
  \institution{University of Washington}
  \city{Seattle}
  \state{Washington}
  \country{USA}
}

\author{Yue Xu}
\email{yuex1@uw.edu}
\affiliation{
  \institution{University of Washington}
  \city{Seattle}
  \state{Washington}
  \country{USA}
}

\author{Devin Murphy}
\email{devinmur@uw.edu}
\affiliation{
  \institution{University of Washington}
  \city{Seattle}
  \state{Washington}
  \country{USA}
}

\author{Sen Zhang}
\email{szhang66@uw.edu}
\affiliation{
  \institution{University of Washington}
  \city{Seattle}
  \state{Washington}
  \country{USA}
}

\author{Te-yen Wu}
\email{tw23l@fsu.edu}
\affiliation{
  \institution{Florida State University}
  \city{Tallahassee}
  \state{Florida}
  \country{USA}
}

\author{Yiyue Luo}
\email{yiyueluo@uw.edu}
\affiliation{
  \institution{University of Washington}
  \city{Seattle}
  \state{Washington}
  \country{USA}
}







\newcommand{\commentText}[1]{#1}    
\newcommand{\red}[1]{\textcolor{red}{#1}}
\newcommand{\TODO}[1]{\commentText{{\color{red}[\textbf{\textsc{TODO}}: \textit{#1}]}}}

\renewcommand{\shortauthors}{Yu, et al.}

\begin{abstract}
Battery-free RFID systems offer a scalable and maintenance-free approach to interaction. We present KnitID, a machine-knitted textile RFID antenna design that enables on-body authentication, localization, and interaction. Unlike prior antenna designs, KnitID achieves a compact antenna form factor (60mm × 8mm) by integrating magnet wire into the unique loop-over-loop structure of machine knitting. This structure reduces the size of conventional loop antennas by around 90\%, while also providing 30\% longer sensing ranges than standard dipole designs with similar size on the human body. The compact form factor creates new opportunities to embed multiple RFID tags across the human body, enriching backscatter signals and supporting a broader range of battery-free on-body interactions. To demonstrate this capability, we build an interactive sleeve to support wearer authentication, spatial localization, and interaction detection. Through technical evaluations, we show the feasibility of KnitID to provide diverse and battery-free interactions on knitted user interfaces.


\end{abstract}



\begin{teaserfigure}
  \includegraphics[width=\textwidth]{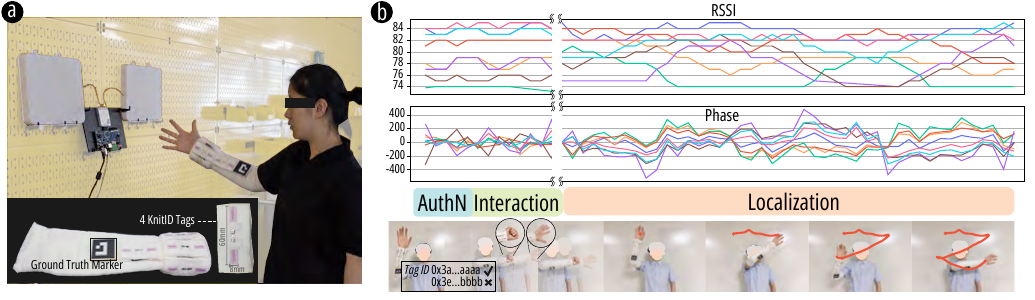}
  \caption{Overview of KnitID. (a) System setup with four RFIDs featuring machine-knitted antennas integrated into a sleeve, and two readers in the environment. (b) Backscattered responses (RSSI and phase) from the readers for authentication, interaction detection (hand clenching), and localization (trajectory recovery).}
  \label{fig:teaser}
\end{teaserfigure}


\maketitle

\section{Introduction}

Battery-free wireless technologies offer a promising pathway toward scalable, unobtrusive, and maintenance-free interactive interfaces. \cite{zada2023battery, xu2025bit, takahashi2024picoring}.
In particular, radio-frequency identification (RFID) systems provide a compelling platform for battery-free interfaces, offering passive communication \cite{liu2021rfid, yang2021rfid}, wireless power harvesting \cite{zada2023battery}, and infrastructure-level connectivity \cite{liu2019tagsheet}. Prior work has explored embedding RFID tags into everyday objects for identification and sensing \cite{li2015idsense, liu2021rfid}, incorporating them on the body for localization and activity recognition \cite{jin2018towards, wang2019rfid}.

RFID tags typically consist of an off-the-shelf chip and an antenna \cite{maia2023using}. In most cases, the antenna dominates the physical form of the tags, largely determining their sizes, mechanical properties and reading ranges.
Existing flexible and wearable RFID tags commonly realize antennas as additional layers attached to a substrate. This lack of seamless integration can lead to detachment over time, limiting reliability for continuous on-body use.

In this work, we present KnitID, a machine-knitted RFID antenna design that support battery-free authentication, interaction and localization. KnitID reduces the size of conventional loop antennas by around 90\% \cite{tsai2013inductively, loop_antenna} while extending sensing range by around 30\% compared with naive knitted dipole designs. Its compact size enables dense integration of multiple RFID tags within a single garment without compromising wearability. To demonstrate its' capability, we build an interactive sleeve that incorporates four KnitIDs for on-body sensing and interaction. Our prototype achieves 100\% authentication accuracy, over 90\% interaction detection accuracy, and a localization Mean Squared Error(MSE) of 5cm.
\section{KnitID and Design Optimization}
KnitID is a compact RFID loop antenna design realized by leveraging the loop-over-It supports three core functionalities: authentication, localization, and interaction detection. All three require an effective antenna with stable gain. Thus, we treat read range and gain as primary objectives and first optimize the antenna design.

\begin{figure}
    \centering
    \includegraphics[width=1\linewidth]{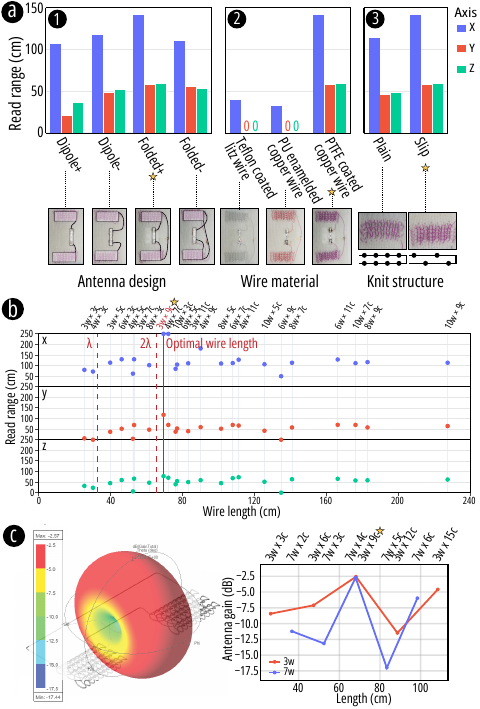}
    \caption{(a) Read range (with default 50cm away from reader for Y/Z axis) of machine-knitted RFID antennas with varying antenna geometry (1), wire materials (2), and knitting structures (3). (b) Read range with varying wale and course. (c) Simulation results of gains.}
    \label{fig:fig2}
\end{figure}

\paragraph{Empirical Characterization}
We fabricate and evaluate antennas across key design parameters, including antenna geometry, conductive materials, knitting structures, wale (\textit{w}), and course (\textit{c}), as shown in Fig. \ref{fig:fig2}a and b. Results show that machine-knitted antennas using folded+, PTFE-coated copper wire, and alternating slip structure achieve the longest read ranges. Among all tested sizes, the 3w $\times$ 9c design performs best, likely because its total conductive length is approximately twice the wavelength. We further validate this observation through simulation.

\paragraph{Modeling and Simulation}
Based on the empirical results, we simulate slip knitted antennas of different sizes by varying \textit{w} and \textit{c}. The results show that antenna gain peaks when the total conductive length approaches 2$\lambda$, validating the trend observed in the empirical characterization.
\section{Multimodal Sensing via KnitID}
To demonstrate the potential of KnitID, we create an interactive sleeve integrated with four KnitID on the cuff (Fig. \ref{fig:teaser}). We also build a system that combine authentication, interaction and localization.

\begin{figure}
  \includegraphics[width=\columnwidth]{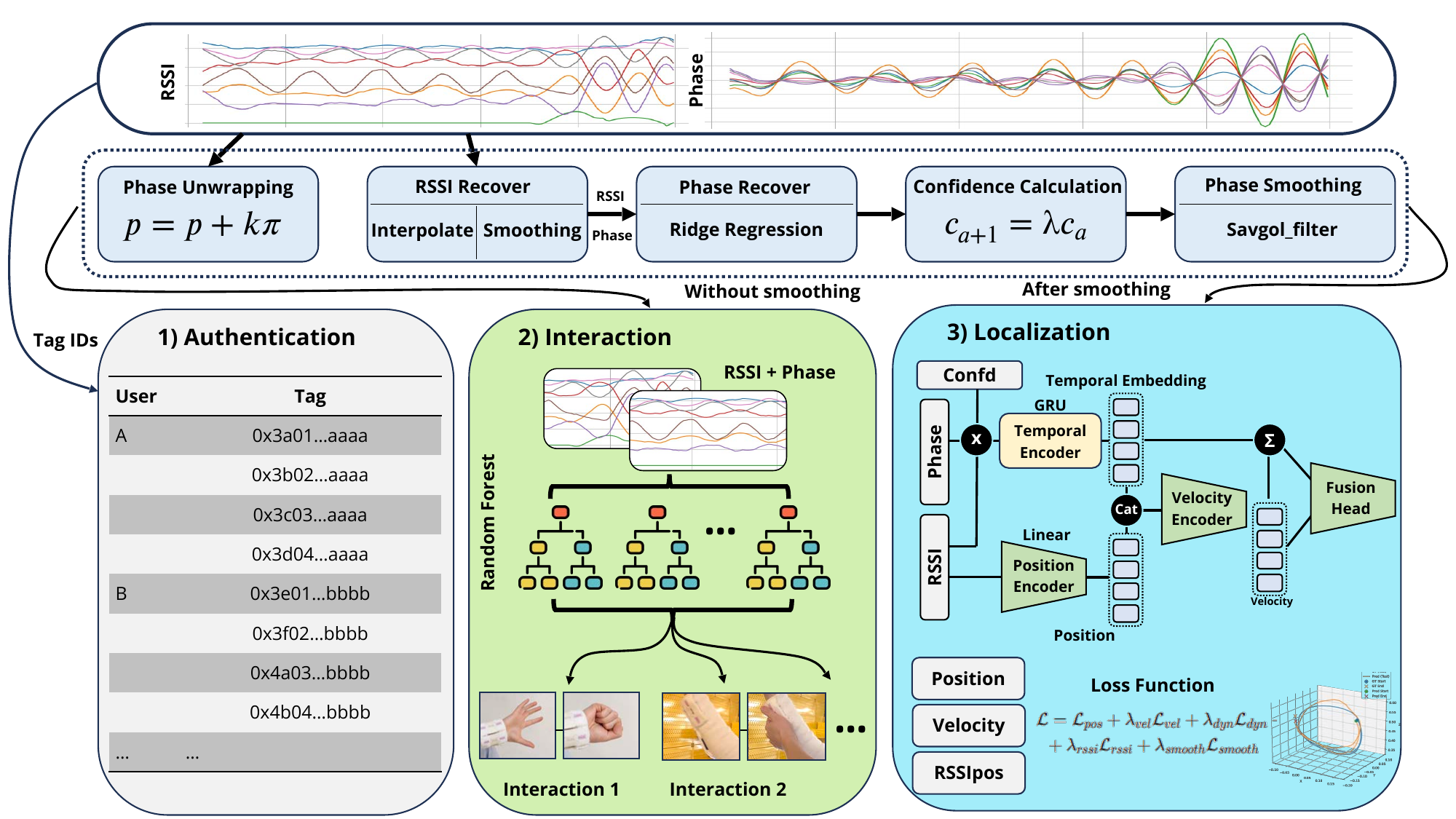}
  \caption{Overview of the multimodal sensing pipeline.}
  \label{fig:pipeline}
\end{figure}

\paragraph{Hardware Prototype}
We fabricate a wearable sleeve integrated with multiple KnitID tags. Specifically, three tags are placed around the wrist, and one additional tag is positioned on the forearm, as illustrated in Fig.~\ref{fig:teaser}. The deployed antennas continuously capture the unique IDs and backscattered signal responses from these tags. We deploy two 9 dBi RFID antennas (\cite{UHF-rfid-antenna}) connected to a reader board with \~15Hz sampling rate based on the Impinj E710 (\cite{readerboard}), mounted on a wall to continuously scan the tags, as shown in Fig.~\ref{fig:teaser}.

\paragraph{Sensing Algorithm}
As shown in Fig. \ref{fig:pipeline}, the system first performs signal preprocessing to improve data quality, including phase unwrapping, RSSI and phase recovery, and confidence estimation. Authentication then verifies the combination of RFID tag IDs, achieving 100\% accuracy across 10 subjects. For interaction recognition, we use RSSI, phase, and confidence features to capture temporal patterns with a random forest, achieving over 90\% accuracy for fist clenching and arm holding. For localization, we use a GRU-based temporal encoder to fuse smoothed RSSI and phase signals. With a loss function combining position and velocity, the model achieves a localization MSE of 5cm.

\paragraph{Applications}
We present a subject-aware drawing board as an example that unifies authentication, interaction, and localization (Fig. \ref{fig:teaser}(b)). Upon entering the sensing area, the system identifies the user via RFID and associates subsequent interactions with that user. A fist-clenching gesture initiates drawing, after which hand trajectories are tracked to render strokes in real time. The system maintains user-specific states and seamlessly switches context when a new user is authenticated, enabling multi-user interaction without explicit switching. In addition to jointly leveraging KnitID’s authentication, interaction, and localization capabilities, the system can also function as a standard RFID-based interface.

\bibliographystyle{ACM-Reference-Format}
\bibliography{references}

\end{document}